\documentclass[sigconf]{acmart}
\usepackage{booktabs} 
\usepackage{lipsum}
\usepackage{tabularx,ragged2e,booktabs}
\usepackage{float}
\usepackage{amssymb}
\usepackage{mathtools}
\usepackage{xcolor}
\usepackage{bbm}
\usepackage{dsfont}
\usepackage{epsfig}
\usepackage{setspace}
\usepackage{subfigure}
\usepackage{url}
\usepackage{hyperref}
\usepackage{microtype}
\usepackage{graphicx}
\usepackage{tabularx}
\usepackage{tabulary}
\usepackage{color}
\usepackage{arydshln}
\usepackage[]{caption}
\usepackage{enumitem}
\usepackage{booktabs} 
\usepackage{lipsum}
\usepackage{tabularx,ragged2e,booktabs}
\usepackage{float}
\usepackage{amssymb}
\usepackage{mathtools}
\usepackage{xcolor}
\usepackage[utf8]{inputenc}
\usepackage[english]{babel}
 \usepackage[most]{tcolorbox}

\usepackage{amsthm}

\setcopyright{acmcopyright}

\newcommand{\specialcell}[2][c]{\begin{tabular}[#1]{@{}c@{}}#2\end{tabular}}

\newcommand{\figsize}{{1\columnwidth}}

\copyrightyear{2018} 
\acmYear{2018} 
\setcopyright{acmcopyright}
\acmConference[CHIIR '18]{2018 Conference on Human Information Interaction \& Retrieval}{March 11--15, 2018}{New Brunswick, NJ, USA}
\acmPrice{15.00}
\acmDOI{10.1145/3176349.3176886}
\acmISBN{978-1-4503-4925-3/18/03}

\begin{document}
\title{Term Relevance Feedback for Contextual Named Entity Retrieval}




\author{Sheikh Muhammad Sarwar, John Foley, and James Allan}
\affiliation{%
  \institution{Center for Intelligent Information Retrieval\\
  College of Information and Computer Sciences\\University of Massachusetts Amherst}
}
\email{{smsarwar,jfoley,allan}@cs.umass.edu}

\begin{abstract}

We address the role of a user in Contextual Named Entity Retrieval (CNER), showing (1) that user identification of important context-bearing terms is superior to automated approaches, and (2) that further gains are possible if the user indicates the relative importance of those terms. CNER is similar in spirit to List Question answering and Entity disambiguation. However, the main focus of CNER is to obtain user feedback for constructing a profile for a class of entities on the fly and use that to retrieve entities from free text. Given a sentence, and an entity selected from that sentence, CNER aims to retrieve sentences that have entities similar to query entity. This paper explores obtaining term relevance feedback and importance weighting from humans in order to improve a CNER system. We report our findings based on the efforts of IR researchers as well as crowdsourced workers.

\end{abstract}

%
%

%
%
%
%

\maketitle
\setlength{\belowcaptionskip}{-1mm}
\section{Introduction}
Entity list retrieval is an important and well motivated problem that has been addressed for more than a decade by the IR community \cite{trecqa05,trecqa06,Demartini:2009,trec2010overview,DalviCC10}. This problem assumes that a user has a well-defined information need that can be expressed using a set of keywords for submitting to an entity ranking system. Some variations allow the user to provide an example entity along with its textual description. The ranking system then returns a list of entities ordered by their relevance to the example entity as well as the user description.

We consider a scenario for list entity retrieval where the entity need is formed on-the-fly, for example, when a user finds an interesting entity in a text excerpt. Small touch-screen devices make this situation more likely as at any time a user can only view a small part of a document, perhaps a sentence or a paragraph. That focused region with an entity of user interest provides contextual clues for a system that would tackle the problem of finding related entities given the example entity.  Part of the challenge is that not every token in this short contextual window is important and some of them might hurt the system's performance by driving the entity retrieval system in the wrong direction. 

As an example, consider a user who reads a document and comes across the sentence: \emph{Carolyn and her twin sisters, Lauren and Lisa, were raised by their mother Ann Freeman, a teacher and administrator in the New York public schools, and their stepfather, orthopedic surgeon Richard Freeman}.
\footnote{Sentence taken from TREC List QA collection that we have used as dataset.} 
Assume that the user wants to know the name of all the family members of Carolyn Bessette, and she pointed out \emph{Ann Freeman} as an example of the target entity class. The context terms related to family members in the given sentence are \textbf{sisters}, \textbf{mother}, and \textbf{stepfather}, and the terms from the entities of interest are \textbf{Lauren}, \textbf{Lisa}, \textbf{Ann}, \textbf{Freeman}, and \textbf{Richard}. Other terms -- such as \textbf{teacher} or \textbf{surgeon} -- might direct the search in a completely different direction; that is, away from family members by focusing on professions rather than family relationships. 



This work addresses the above problem, and proposes context term selection using two approaches: (1) top-k keyword extraction from a sentence based on keyword and example entity similarity, and (2) weighted term relevance feedback (TRF) from users. We focus on the latter and use the former automated process as a baseline as we explore the following research questions: 

\newtheorem*{remark}{RQ1}
\begin{remark}
Does a user's term-level relevance feedback provide 
improved  results for the CNER task in comparison to fully automated baselines?

\end{remark}


\newtheorem*{remark1}{RQ2}
\begin{remark1}
Is the user feedback more effective for CNER if the user can indicate which of the terms is more important? 
\end{remark1}


Given a starting sentence and an identified entity, the output of CNER is a ranked list of sentences that are most likely to contain an instance of the desired entity class (as inferred from the query sentence). We impose a novelty requirement, also, such that a sentence is only relevant if it includes at least one relevant entity that has not already been seen in the ranked list.  

We show that user feedback provides a 10.7\% improvement in mAP over strong baselines and that there is an improvement of 14.9\% if the user can provide weights on the selected terms.


\section{Related Work} 

CNER is broadly similar to tasks such as List QA, Entity Ranking, and list completion \cite{trecqa05,trecqa06,Demartini:2009,trec2010overview,DalviCC10}. All of these tasks rely heavily upon external sources of information like Knowledge Bases (KB) to locate an entity of interest and then retrieve similar types of entities based on the contextual evidence present in a question or search query. CNER differs from these tasks as it is focused on entities that are not popular enough to be found in a KB, requiring techniques based solely on the original source text. Moreover, typical solutions to the broad set of tasks do not allow for user interactivity, a specific goal for CNER.

There have been numerous attempts to use term feedback to improve retrieval tasks, but it can be quite a challenge to do so. 
Even though users desire to provide and control the set of terms for query expansion, in most cases it does not lead to better performance \cite{Koenemann:1996rf,Nemeth:2004,Ruthven:2003}.
Kelly et al. showed that if a sentence is provided as an example use of an expansion term, users can slightly improve the precision of the retrieval system in comparison with a strong Pseudo Relevance Feedback (PRF) baseline \cite{Kelly:2006}.
Studies suggest that even if the improvement is not great, the availability of interactive TRF is considered a positive aspect of a system by users~\cite{Nemeth:2004,Belkin:2001}. We are inspired from the application of TRF on ad-hoc IR, and incorporated it as a context refinement technique for CNER.   


Our problem is closely related to entity disambiguation that primarily aims to link an entity to a KB given a document about that entity \cite{Sun:2015:MMC}. Entity disambiguation has a different goal 
and we also assume that knowledge about a target entity might not be available in a structured form. Nevertheless, our challenge of finding related sentences could be useful for entity disambiguation. 





\section{Methodology}
We discuss our method of obtaining term relevance feedback from a user for formulating a CNER query, and how we use those terms to obtain a better representation of the query.  

    
    
    



\subsection{Term Relevance Feedback Acquisition}
\subsubsection{Collection}
We use a pool of queries from the TREC 2005 and 2006 List Question Answering (QA) datasets, where relevant entities relevant to each of the list questions are annotated by TREC assessors in several relevant documents. We combine those documents as well as other non-relevant documents from the dataset and break them into sentences. For each list question we then select a sentence that contains at least one entity relevant for that question. The selected sentence and entity becomes a CNER query seeking to find the remaining sentences that contain other relevant entities. Figure~\ref{fig:example} shows two sample list questions and corresponding sentences that we selected from our dataset.  

We selected 20 queries for which there are  nine relevant entities (excluding the query entity) on an average and all sentences that contain those entities. We created two subsets of ten sentence queries and an interface that presents 10 queries to the user one by one (randomized for each user). The interface allows the user to select terms for each query-sentence pair (as described below). We chose 10 queries for a single session because we measured that on an average it takes around 12 minutes for a user to annotate 10 queries. We did this to ensure that users would not be overwhelmed by the length of the task, inspired by the 15 minute attention span limit often cited in education~\cite{wankat2002effective}.

\subsubsection{Interface for Query Generation}
Each user was asked to look at a sentence containing an entity and then select the words within that sentence that seemed most likely to be useful as query words if someone wanted to find other examples of entities of the same type. The user was also asked to indicate relative importance of words by selecting the best words more often.

\begin{figure}[t]
\begin{tcolorbox}
	\includegraphics[width=\figsize]{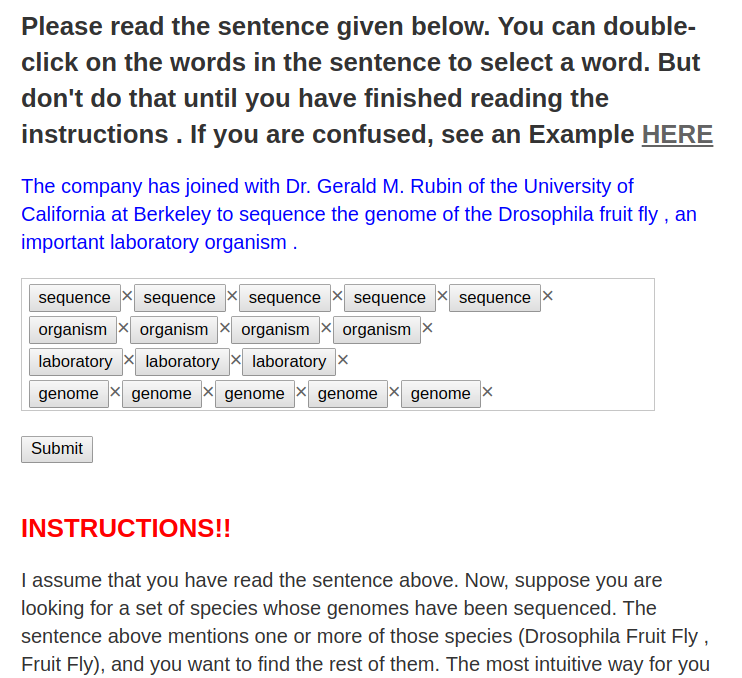}
\end{tcolorbox}
\caption{Interface for Obtaining Contextual Keywords} \label{fig:interface}
	
\end{figure}

Figure~\ref{fig:interface} shows a screen-shot of the interface that we used to obtain this term relevance feedback. The interface does not return any search result, but it facilitates users' providing terms and their importance by double-clicking them. In the Figure, clicked words are shown in small grey boxes. More important words were clicked more often and appear multiple times. After the annotation process we processed the collected queries and all the baselines offline.


\begin{figure}[t]
\begin{tcolorbox}
\flushleft
\textit{``Who testified in defense of Susan McDougal?''} \\
$\rightarrow$ ``Susan McDougal 's lawyer says she plans to attend the opening of Independent Counsel Kenneth Starr's latest case, the trial of a Virginia woman who provided helpful testimony for Mrs. McDougal a month ago.''

\textit{``Programs sponsored by the Lions Club [International]''} \\  $\rightarrow$ ``Lions Club International, the world's largest service club organization, plans to help more cataract sufferers in China as an example of blindness prevention for the rest of the world, according to Tam Wing Kun, chairman of the Sight First China Action (SFCA) project.''
\vspace{0.5mm}
\end{tcolorbox}
\caption{Example of List Question, and a corresponding sentence for annotation are presented below. }
\label{fig:example}

\end{figure} 

\subsubsection{Subjects}
We collected context word annotations for 20 queries. For each query we took annotation from six Mechanical Turk users from the United States and three lab researchers familiar with Information Retrieval and thus likely to be better at selecting terms and judging their relative importance. 

\subsection{Retrieval Method}
\label{sec:sentence_embedding}
Vocabulary mismatch is a problem that particularly affects short-text retrieval and semantic features are an effective way to alleviate this problem. Our retrieval task also demands the use of semantic similarity because we do not seek to find entities that appear exactly as described in the query sentence. We combine the effectiveness of both syntactic and semantic matching for computing sentence similarity.  We use BM25 for keyword matching and Sentence Embedding (SE) ~\cite{WietingBGL15a} to compute semantic similarity. We assume that word matching is more important for the user selected words and semantic similarity is important for matching similar entities, and combine the benefits of both to create an effective model. 

We capture semantic level matching between a sentence pair to retrieve and score the top $k$ sentences against the query sentence. We use the average of word embedding to obtain sentence embedding for the query and candidate sentences. Word embedding methods learn a low-dimensional vector representation of words from a large, unstructured text corpus; we use the skip-gram model proposed by Mikolov et al.~\cite{mikolov2013efficient} to generate representations for words. Finally, we use cosine similarity to compute similarity between a query and a candidate sentence. Our approach is inspired by Wieting et al.~\cite{WietingBGL15a}, who showed that a simple averaging over the embedding of the words in a sentence provides an effective representation for that sentence and that representation is particularly helpful for sentence similarity task.

For all our ranking techniques, we use the Stanford Named Entity Recognizer~\cite{Finkel:2005} to reject candidate sentences that do not contain entities of the appropriate type. While this may introduce false-negatives, it greatly increases precision of our system, and allows our other techniques to focus on ranking and increasing recall.

%

\subsection{Query Expansion (QE)}
\label{sec:QE}
In order to obtain a broader and generalized representation of query sentence, we use Pseudo Relevance Feedback (PRF) for query expansion at the sentence level. 
We use BM25 to retrieve PRF sentences given the query sentence and compute the average over the embedding of those sentences to obtain a more robust representation of the query. 

\subsection{PRF with User Feedback}
\label{sec:prf_bm25}
We make use of the context words selected by user for finding PRF sentences with BM25 technique. We expect that a keyword-based search technique would find sentences focusing on user selected terms. Suppose, our original query sentence $Q_o$ contains a list of $n$ terms, and user $u$ has constructed a list, $CW$ from $Q_o$, of $k$ words, where each word appears there one or more times. Now, in order to get an expanded query $Q_e$, we simply concatenate all the terms in $CW$ with $Q_o$. The goal of this process is to assign term importance in a query by repeating the term multiple times. Even though it is not a sophisticated method of incorporating user-provided term weights, it works well in practice. We search the sentence corpus with $Q_e$ and use the top $k$ retrieved sentences, $S_{\mbox{\it topk}}$  for obtaining a better representation for $Q_o$. Finally, we compute the average of the sentence embeddings from the sentences in set $Q_f = \{Q_o \cup S_{\mbox{\it topk}}\}$, perform SE based search using $Q_f$ and re-rank them using the method described in \ref{sec:sentence_embedding}.

\section{Experimental Results}

\setlength\dashlinedash{0.5pt}
\setlength\dashlinegap{1.5pt}
\setlength\arrayrulewidth{0.5pt}

\begin{table*}[!ht]
\centering
\caption{Average performance of various methods. {\rm Measures are listed in the first row, with high-precision measures listed first. mAP and R are cut-off at depth 1000. The first section of the table presents baselines, then weighted feedback and finally unweighted feedback. The percentage improvement is shown over the Sentence Embedding (SE) baseline.}}
\label{tab:batch_average}
\begin{tabular}{|c|c|c|c|c|c|l|l|l|}
\hline
R@5   & R@10  & P@5   & P@10  & mAP & R & Method              & Source of Context Words  & Weighted? \\ \hline \hline
0.145 & 0.234 & 0.180  & 0.180  & 0.188    & 0.891  & SE                  & None                     & No      \\
0.123 & 0.183 & 0.160  & 0.130  & 0.153    & 0.884  & SE + PRF     & None                     & No      \\
0.147 & 0.177 & 0.200   & 0.150  & 0.162    & 0.865  & SE + PRF + CW & Sim (Entity, Word)       & No        \\ \hline
\specialcell{0.183\\(+26.3\%)} & \specialcell{0.244\\ (+4.3\%)} & \specialcell{0.231\\ (+28.4\%)} & \specialcell{0.184 \\(+2.3\%)} & 
\specialcell{0.216 \\ (+14.9\%)} &
\specialcell{0.910\\(+2.2\%)}   & SE + PRF + CW & Mturk + Lab Participants & Yes \\ \hdashline
\specialcell{0.204\\ (+40.7\%)} & \specialcell{0.267\\ (+14.2\%)} & 
\specialcell{0.237\\ (31.7\%)} & \specialcell{0.196\\ (+8.9\%)} & 
\specialcell{0.232\\ (+23.5\%)}    & \specialcell{0.921\\ (+3.4\%)}  & SE + PRF + CW & Lab Participants         & Yes       \\ \hdashline
\specialcell{0.171\\(+18\%)} &
\specialcell{0.229\\(-2.2\%)} & 
\specialcell{0.222 \\(+23.4\%)} & 
\specialcell{0.176\\ (-2.3\%)} & 
\specialcell{0.205\\ (+9.1\%)} & 
\specialcell{0.900\\ (+1.1\%)} & SE + PRF + CW & Mturk Participants       & Yes       \\ \hline
\specialcell{0.175\\(+20.7\%)} & 
\specialcell{0.234\\(+0.0 \%)} & 
\specialcell{0.226 \\ (+25.6\%)} & 
\specialcell{0.179 \\ (-0.6\%)} & 
\specialcell{0.208 \\ (+10.7\%)} & 
\specialcell{0.907 \\ (+1.1\%)} & SE + PRF + CW & Mturk + Lab Participants & No        \\ \hdashline
\specialcell{0.186\\(+28.3\%)} & 
\specialcell{0.255 \\(+9.0\%)} & 
\specialcell{0.234 \\ (+30\%)} &
\specialcell{0.194 \\ (+7.8\%)} & 
\specialcell{0.220 \\ (+17.1\%)} &
\specialcell{0.920 \\ (+3.3\%)}
& SE + PRF + CW & Lab Participants         & No \\ \hdashline 
\specialcell{0.166\\(+14.5\%)} &
\specialcell{0.219\\(-6.5\%)} & 
\specialcell{0.217\\(+20.6\%)} & 
\specialcell{0.168\\(-6.7\%)} & 
\specialcell{0.199\\(+5.9\%)} &
\specialcell{0.897\\(+0.7 \%)}
& SE + PRF + CW & Mturk Participants       & No        \\ \hline
\end{tabular}
\vspace{-2mm}
\end{table*}

\subsection{Evaluation Metrics and Relevance}

We use novelty versions of recall and precision, standard measures modified so that only the first instance of a target entity is considered relevant. We use recall@$k$ to measure the number of relevant (and unique) entities observed in the top $k$ sentences and we use precision@$k$ to measure the proportion of sentences in the top $k$ that contain relevant (and unique) entities. We stress that the relevance of a sentence is determined by two properties: containing a relevant entity \emph{and} being unique in the ranked list so far. We also report MAP@1000 and recall@1000, measures that are important because when we want to perform two-stage retrieval and ranking, retrieving most of the entities in the top 1000 sentences becomes crucial.


\subsection{Baseline Methods}
We compare the effectiveness of user feedback against three non-interactive baselines.
\begin{itemize}[leftmargin=*]
\item \textbf{SE} is the Sentence Embedding (SE) based search described in Section~\ref{sec:sentence_embedding} that assumes no information regarding term importance. 

\item \textbf{SE + PRF} is similar to the approach described in Section~\ref{sec:prf_bm25} that uses BM25 to retrieve the top $k$ sentences using the sentence query and then combines those to obtain an expanded query that is used with SE. 

\item \textbf{SE + PRF + CW + Sim (Entity, Token)} is similar to the process of integrating user-selected context words into the query as mentioned in Section~\ref{sec:prf_bm25}. However, the process of obtaining context word is not based on any human input. We use this baseline to check how better human input is compared to an automatic process that can generate context words. This method computes the similarity of each word in the query sentence with the query entity. Then it uses the five most similar words for performing PRF. Similarity between a word and query entity is computed using the similarity of their embedding. An entity embedding is constructed by the average of the embedding of the words in it. 

\end{itemize}


\subsection{Result Discussion}
Table \ref{tab:batch_average} summarizes the average performance of term relevance feedback for CNER across the queries that have been annotated by lab and Mechanical Turk participants. Across all forms of feedback, the lab participants created more effective queries than the crowdsource workers: this is reasonable as the lab participants are likely to be more expert searchers. 

Overall, term feedback was helpful (10.7\% improvement in mAP), and weighted term feedback was even more helpful (14.9\% improvement in mAP). The means that our two research questions are both answered positively: user feedback provides improved results for CNER, and allowing users to specify an ordering or weighting on terms is helpful.

\begin{figure}[t]
\includegraphics[width=\figsize]{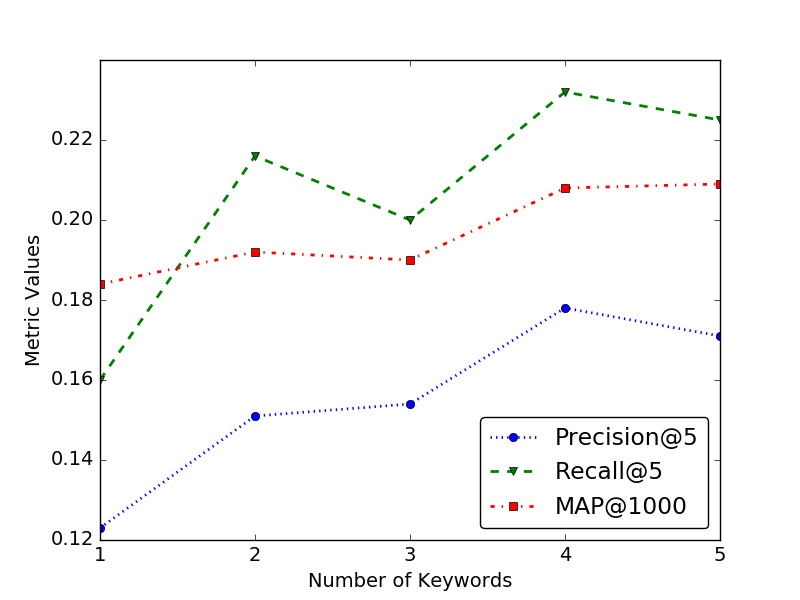}
\caption{Performance Sensitivity with Keywords Addition} 
\label{fig:keyword_effect}
\vspace{-5mm}
\end{figure}

In addition, we analyze the impact of adding fewer keywords (and therefore minimizing user involvement). For each of the queries, we selected the top-$k=1\ldots5$ terms based on the weights provided by the users, added them to the original query (with weights). Performance is presented in Figure~\ref{fig:keyword_effect} in terms of Precision@5, Recall@5 and mAP@1000. Although there is some noise, particularly in the recall of the solution, it is clear that a handful of keywords can be effective (although it does depend on the user and the quality of the terms selected), but more terms do appear to be better.

\section{Conclusion}
We adopt a term relevance feedback technique for list query construction from a sentence and show its effectiveness in entity retrieval. We started this work with two research questions and answered them both affirmatively. We showed that (RQ1) users \emph{can} select better query terms than automatic methods, and that (RQ2) it \emph{is} helpful for the user to identify which terms are best. Our interface for collecting this information was rudimentary and we did not explore alternatives for this study. Future work will look at how an interface can best support a user in providing that information.
\vspace{-2mm}
\section*{Acknowledgement}
This work was supported in part by the Center for Intelligent Information Retrieval and in part by NSF grant \#IIS-1617408. Any opinions, findings and conclusions or recommendations expressed in this material are those of the authors and do not necessarily reflect those of the sponsors. 
\bibliographystyle{acm}
\bibliography{sigproc}

\end{document}